\documentclass[global,twocolumn]{svjour}
\usepackage{textcomp}
\usepackage{graphics}
 \usepackage{epstopdf}
\usepackage{braket}
\usepackage{mathtools}
\usepackage{amsmath}
\usepackage{subfigure}
\usepackage{subfig}
\usepackage{xcolor}
\usepackage{multirow}
\usepackage{cite}
\usepackage{changes}
\usepackage{ulem}
\usepackage{selinput}
\usepackage[english]{babel} 
\journalname{Appl Phys B}
\begin{document}
\title{Holographic method for site-resolved detection of a 2D array of ultracold atoms }
\author{Daniel Kai Hoffmann$^1$, Benjamin Deissler$^{1,2}$, Wolfgang Limmer$^1$, and Johannes Hecker Denschlag$^1$}
%
\institute{$^1$ Institut f\"ur Quantenmaterie, Universit\"at Ulm, 89081 Ulm, Germany\\$^2$ Leica Mikrosysteme Vertrieb GmbH, 35578 Wetzlar, Germany\\Corresponding author: johannes.denschlag@uni-ulm.de}
\date{\today}
\authorrunning{D. K. Hoffmann et al.}
\titlerunning{Site-resolved holographic detection of ultracold atoms}
\maketitle
\begin{abstract}We propose a novel approach to site-resolved detection of a 2D gas of ultracold atoms in an optical lattice. A near
resonant laser beam is coherently scattered by the atomic array and its interference pattern is holographically recorded
by superimposing it with a reference laser beam on a CCD chip. Fourier transformation of the recorded intensity pattern
reconstructs the atomic distribution in the lattice with single-site resolution. The holographic detection method
requires only a few hundred scattered photons per atom in order to achieve a high reconstruction fidelity.
Therefore, additional cooling during detection might not be necessary even for light atomic elements such as lithium.
\end{abstract}
\section{Introduction}
\label{intro}
Ultracold atoms in optical lattices allow for investigating many-body physics in a very controlled way
(see e.g. \cite{IB08}). For such experiments site-resolved detection of the exact atomic distribution in the lattice
can be very advantageous and it has  
recently been demonstrated \cite{Weiss07,Bak09,She10,Zw15,Kuh15,MG15}.
In these experiments, the fluorescence of illuminated atoms is detected using a high-resolution objective.
{During the imaging process, typically several thousand photons are scattered per atom. This leads to strong
heating of the atoms, requiring additional cooling.}\\
Alternative imaging techniques using the diffraction of a laser beam by an atomic ensemble have been demonstrated for the
detection of cold atomic clouds \cite{Tur04,Tur05,And96,Lui13,Wu14,Kad01}. However, these techniques have neither been discussed for
single-particle resolution nor single-site detection.

Here, we propose to image an atomic array with high resolution by using a variation of the off-axis holography technique of Leith and
Upatnieks \cite{Kad01,LU62}. Two coherent laser beams are used to record the hologram of an illuminated atomic array. One acts
as a probe beam and is coherently scattered by the atoms \cite{Wei11}, while the other acts as a reference beam which bypasses the atoms. Both
beams are superimposed to interfere and to generate the hologram which is recorded with a charge-coupled device (CCD) camera.
An algorithm based on fast Fourier transformation reconstructs an image of the atomic array. The reference beam fulfills two
purposes: On the one hand it separates the holographic image from disturbing low spatial frequency signals in the reconstruction.
On the other hand it strongly amplifies the atomic signal, as in spatial heterodyne detection \cite{Kad01}. This allows the use of a weak probe beam
while keeping the signal high compared to detection noise.
We estimate that for our scheme the number of scattered photons per atom can be small enough ($\approx$ 150 photons) such that single site detection could be realized without additional cooling.
Moreover, the scheme might open the path for multi-particle detection per lattice site, since the low photon flux reduces 
 photoassociation.

The paper is organized as follows: Section~\ref{detection_scheme} sketches the basic scheme of the holographic detection method.
Section~\ref{theoretical_description} reviews theoretical background on atom light interaction, and on optical signals. In Section~\ref{simulations}, we pre\-sent the results of
numerical calculations for the concrete example of $^6\textrm{Li}$ atoms in an optical lattice. Furthermore,
 we discuss the conditions for which a successful reconstruction of an atomic distribution can be achieved.
Section~\ref{conclusion} concludes with a short summary and an outlook.

\section{Detection scheme\label{detection_scheme} }

\begin{figure}[!h]
\includegraphics{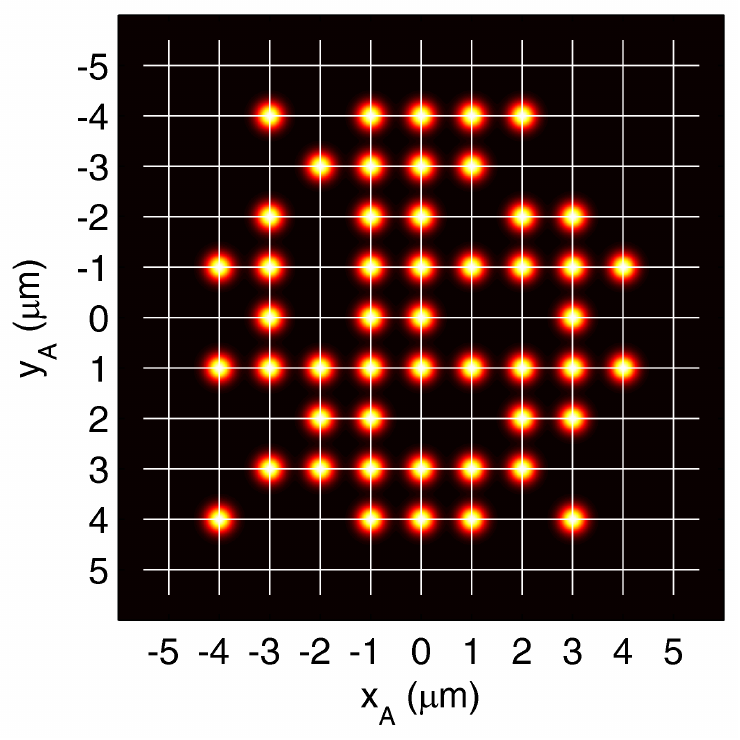}
\caption{Ensemble of $N_\mathrm{A}$ = 50 atoms, distributed over a 2D square lattice with lattice constant $a=1\,\mu$m.}
\label{fig:lattice}
\end{figure}

We discuss the proposal in terms of a concrete example. As depicted in Fig.~\ref{fig:lattice}, we consider an ensemble of
$N_\mathrm{A}$ = 50 atoms distributed over a 2D lattice with $11 \times 11$ sites and a lattice constant of $a=1$\,$\mu$m.
Each site is either empty or occupied by one single atom. We assume the lattice potential to be deep enough such that tunneling between the
lattice sites is negligible. 

\begin{figure}[ht]
\includegraphics{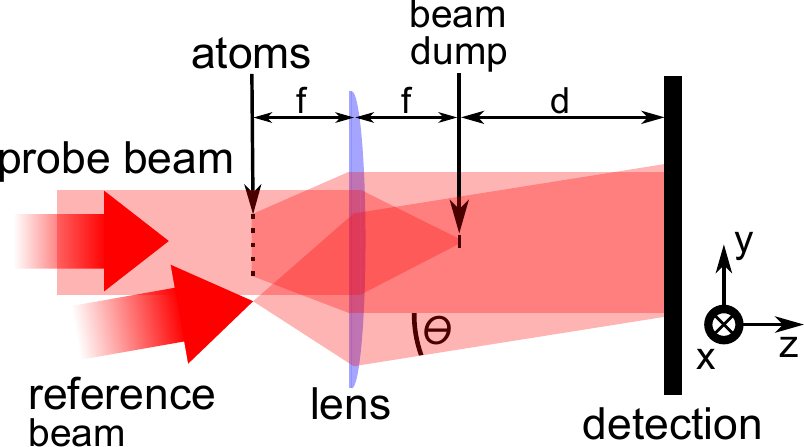}\\
\caption{Basic scheme of the holographic detection method. A probe beam illuminates the 2D array of atoms
and the scattered light is collimated by a lens with focal length $f$. The scattered light is superimposed
with a reference beam on the CCD detector which is placed at a distance $d$ behind the back focal plane.
A beam dump blocks the unscattered light.}
\label{fig:scheme0}
\end{figure}

The overall setup for the  detection method is shown in Fig.~\ref{fig:scheme0}. A Gaussian laser beam,
near resonant to an optical atomic transition, is split into two beams, the probe and the reference beam.
The probe beam propagates perpendicularly to the atomic layer and illuminates the atoms in the
optical lattice. It has a diameter much larger than the spatial extent of the atomic sample, such that its electric field strength
is approximately the same for all atoms.

The atoms are treated as Hertzian dipoles that coherently scatter the probe light. The scattered light is collimated
by a diffraction-limited lens with a large numerical aperture and forms nearly perfect plane waves, which propagate towards the CCD detector.
Since the spatial extent of the atomic sample is typically about three orders of magnitude
smaller than the focal length $f$ (Fig.~\ref{fig:scheme0} is not to scale!), the plane waves propagate almost parallel to the optical
axis along the $z$ direction.

The non-scattered part of the probe beam is blocked by a small beam dump in the back focal plane of the lens.
The reference beam bypasses the atomic layer and is superimposed with the collimated scattered light in the detection plane at an
angle $\theta$. In order to keep $\theta$ small (see discussion in Sec.~\ref{simulations}), the reference beam is transmitted through
the same lens as the scattered probe light. For this purpose, it is strongly focussed to a micrometer spot size in the front focal
plane (at a sufficiently large distance to the atoms) and then collimated by the lens.

The overall intensity pattern is recorded by a CCD camera with a high dynamic range in order to resolve weak interference
fringes on a high background signal. The pattern is subjected to a 2D Fourier transform (FT) \cite{Jue02} which directly
yields the atomic distribution in the lattice. This step is analogous to classical holography where a readout wave reconstructs the original object, corresponding to the holograms Fourier transform.\cite{Goo05}.

\section{Theoretical description}
\label{theoretical_description}

\subsection{Coherent light scattering}

We use a semi-classical model for the interaction of a single atom with a monochromatic coherent light field. Each atom acts
as a quantum mechanical two-level system with transition frequency $\omega_0$. The atom is driven by a weak external laser field
with frequency $\omega$. This leads to photon scattering with a rate \cite{Lou09}
\begin{equation}
R_\mathrm{S} = \frac{\Gamma}{2}\frac{I/I_\mathrm{sat}}{1 + \left(2\Delta/\Gamma\right)^2 + I/I_\mathrm{sat}}
= \frac{N_\mathrm{ph}}{T_\mathrm{ac}},
\label{eq:Rs}
\end{equation}
where $I$ denotes the incident intensity of the driving field, $I_\mathrm{sat}$ the saturation intensity of the
atomic transition, and $\Delta = \omega - \omega_0$ the detuning between laser and transition frequency. $\Gamma $  is
the linewidth of the atomic transition, and $N_\mathrm{ph}$ the total number of scattered photons per atom within the acquisition
time $T_\mathrm{ac}$.

In general, the intensity $I_{\mathrm{sc}}$ of the scattered light consists of both coherently and
incoherently scattered parts. The coherent fraction of the scattered light  $I_\mathrm{coh}/I_{\mathrm{sc}}$ is given by \cite{Mol69,Lou09}
\begin{equation}
\frac{{I}_\mathrm{coh}}{{I_{\mathrm{sc}}}} = \frac{1+ \left(2\Delta/\Gamma\right)^2}{1 + \left(2\Delta/\Gamma\right)^2 + I/I_\mathrm{sat} }.
\label{eq:Icoh}
\end{equation}
A weak incident beam with large detuning will therefore yield mainly coherently scattered light.
As a concrete example, choosing $I/I_\mathrm{sat}< 1$ and $\Delta = - \Gamma$ yields mostly coherent emission.

The probe beam as well as the  reference beam ($\theta \approx 1^{\circ}, \phi=45^{\circ}$ see Eq.~(\ref{eq:k_R})) are linearly polarized along the $y$
direction. Treating the atoms as Hertzian dipoles, the electric field at position $\mathbf{r }= (x, y, z)$ in the far field,
emitted by a single atom $n$ at position $\mathbf{r}_n = (x_n,y_n,0)$, is given by
\begin{equation}
 E_{\mathrm{A}}(\mathbf{r},\mathbf{r}_n) = E_\mathrm{A0} \frac{\sqrt{(x-x_n)^2+z^2}}{k\left|\mathbf{r}-\mathbf{r}_n\right|^2}
  e^{ik|\mathbf{r}-\mathbf{r}_n|}\,,
 \label{eq:CoherSc}
\end{equation}
with the wavenumber $k = 2\pi/\lambda$. Integrating the corresponding intensity over the entire solid angle $4\pi$ relates
$E_\mathrm{A0}$ and the total number $N_\mathrm{ph}$ of scattered photons per atom
\begin{equation}
E_\mathrm{A0}^2 = \frac{3 k^2 \hbar \omega}{4\pi c \epsilon_0 T_\mathrm{ac}}N_\mathrm{ph}.
\label{eq:Nph2EA0}
\end{equation}
Here, $c$ denotes the speed of light in vacuum and $\epsilon_0$ the permittivity of free space.

The wave emitted by the $n$th atom in the optical lattice is converted by the lens into a nearly perfect plane
wave with wave vector
\begin{equation}
\mathbf{k}_n = \left( \begin{array}{c} k_{n,x} \\ k_{n,y} \\ k_{n,z} \end{array} \right) =
\frac{k}{\sqrt{x_n^2 + y_n^2 + f^2}} \left( \begin{array}{c} -x_n \\ -y_n \\ f \end{array} \right).
 \label{eq:WaveVector_n}
\end{equation}
The field distribution of the plane wave in the detector plane $z=z_D$ reads
\begin{equation}
 E_{\mathrm{S},n}(x,y) =  E_\mathrm{A0} \, g_\mathrm{A}(x,y)\,
 e^{i(x k_{n,x} + y k_{n,y} + \varphi_n)},
 \label{eq:PlaneWave_P}
\end{equation}
where $\varphi_n$ includes the constant term $z_D k_{n,z}$ and the phase shift acquired by the wave while passing through the
lens.

The field envelope $g_\mathrm{A}(x,y)$ is a slowly varying function which can be determined from  Eq.~(\ref{eq:CoherSc}).
Since the plane waves propagate almost parallel to the z axis behind the lens, $g_\mathrm{A}(x,y)$ is essentially independent of z.
Therefore, we calculate $g_\mathrm{A}(x,y)$ at the position of the lens. Setting $z=f$ in Eq.~(\ref{eq:CoherSc}) and
using the relation $|\mathbf{r}_n| \ll |\mathbf{r}|$ we obtain
\begin{equation}
 g_\mathrm{A}(x,y) \approx \frac{\sqrt{x^2+f^2}}{k(x^2+y^2+f^2)}\Theta(r_{l}-\sqrt{x^2+y^2}).
 \label{eq:gA}
\end{equation}
The Heaviside function $\Theta$ accounts for the finite size of the lens with radius $r_{l}$. 

The electric field of the Gaussian-shaped reference beam at the detector reads
\begin{equation}
 E_\mathrm{R}(x,y) = E_{\mathrm{R0}}\ g_\mathrm{R}(x,y)
 e^{i(x k_{R,x} + y k_{R,y} + \varphi_R)},
\label{eq:reference}
\end{equation}
with the wave vector
\begin{equation}
\mathbf{k}_R = \left( \begin{array}{c} k_{R,x} \\ k_{R,y} \\ k_{R,z} \end{array} \right) =
 k \left( \begin{array}{c} \sin\theta \cos\phi \\ \sin\theta \sin\phi \\ \cos\theta \end{array} \right).
\label{eq:k_R}
\end{equation}
For small $\theta$, the Gaussian field envelope $g_R(x,y)$ is given by
\begin{equation}
g_R(x,y) \approx e^{-(x^2+y^2)/w^2}\Theta(r_{l}-\sqrt{x^2+y^2}),
\label{eq:gR}
\end{equation}
with reference beam waist $w$.

\subsection{Interference and Fourier transformation}

\begin{figure}[ht]
\includegraphics{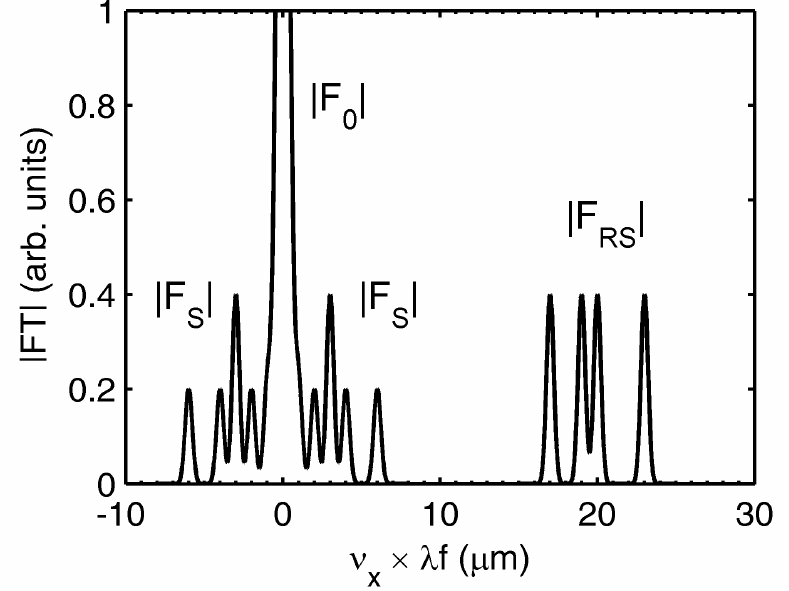}
\caption{1D cut through a schematic 2D FT along the spatial frequency axis $\nu_x$ at $\nu_y = 0$, illustrating the contributions
of $F_\mathrm{0}, F_\mathrm{S}$, and $F_\mathrm{RS}$. The four peaks around $\nu_x \times \lambda f \approx$ 20~$\mu$m reconstruct
the positioning of the four atoms in Fig.~\ref{fig:lattice} arranged along the $x_\mathrm{A}$ axis at $y_\mathrm{A} = 0$.}
\label{fig:FFTschematic}
\end{figure}

The total electric field in the detector plane is obtained by adding up all individual fields. The corresponding intensity,
\begin{equation}
I_\mathrm{D}(x,y) = \frac{1}{2}c\epsilon_0 \left| E_\mathrm{R}(x,y) + \sum_{n} E_{\mathrm{S},n}(x,y)\right|^2 ,
\label{eq:TotalField}
\end{equation}
can be written as a sum of three contributions
\begin{equation}
I_\mathrm{D} = I_0 + I_\mathrm{S} + I_\mathrm{RS}.
\label{eq:TotalIntensity}
\end{equation}
The$\!$$\:$ particle distribution is derived from the Fourier transform $F_\mathrm{D}$ of the intensity profile $I_D$.
$F_\mathrm{D}$ decomposes into three  parts,
$F_\mathrm{0}, F_\mathrm{S}$, and $F_\mathrm{RS}$. This is illustrated in the schematic plot in Fig.~\ref{fig:FFTschematic}, which
depicts a 1D cut through a 2D FT along the spatial frequency axis $\nu_x$ at $\nu_y = 0$. The illustration is consistent with
the atomic distribution in Fig.~\ref{fig:lattice} and presumes a wave vector $\mathbf{k}_R$ with $k_{R,y} = 0$.

The first contribution $I_0$ in Eq.~(\ref{eq:TotalIntensity}) is a broad structureless intensity background
\begin{equation}
I_0 \propto E_\mathrm{R0}^2 \ g^2_\mathrm{R}(x,y) + N_\mathrm{A} E_\mathrm{A0}^2 \ g_\mathrm{A}^2(x,y)
\label{eq:I_0}
\end{equation}
whose FT $F_\mathrm{0}$ is represented by the large peak at the origin in Fig.~\ref{fig:FFTschematic}.
The width of the peak is determined by the inverse beam sizes $g_R$ and $g_A$. The second contribution
\begin{equation}
I_\mathrm{S} \propto E_\mathrm{A0}^2 \ g_\mathrm{A}^2(x,y) \sum_{n>m} \cos [2\pi (\nu_{nmx} x + \nu_{nmy} y) + \Delta\varphi_{nm}]
\label{eq:I_P}
\end{equation}
with $\Delta\varphi_{nm} = \varphi_{n} - \varphi_{m}$ arises from the interference between the electric fields $E_{\mathrm{S},n}$
emitted by the individual atoms in the optical lattice. Since $f \gg x_n, y_n$, the spatial frequencies $\nu_{nmx}$
and $\nu_{nmy}$ are approximately given by
\begin{equation}
\nu_{nmx} = \frac{x_m - x_n}{f\lambda}\,,\,\,
\nu_{nmy} = \frac{y_m - y_n}{f\lambda},
\label{eq:nunm}
\end{equation}
where $(x_n,y_n)$ are the atomic positions in the optical lattice. Each pair of spatial frequencies  $\nu_{nmx},  \nu_{nmy}$
gives rise to a well-defined peak in the FT close to the origin. The width of the peaks is again determined by the inverse of
the collimated beam width $g_A$. The third term in Eq.~(\ref{eq:TotalIntensity}),
\begin{eqnarray}
I_\mathrm{RS} & \propto & E_\mathrm{R0}E_\mathrm{A0} \ g_\mathrm{R}(x,y)\, g_\mathrm{A}(x,y) \times \nonumber \\
& \times & \sum_{n} \cos [2\pi (\nu_{nx} x + \nu_{ny} y) + \Delta\varphi_{nR}],
\label{eq:I_RP}
\end{eqnarray}
arises from the interference of the scattered beams with the reference beam. Here,  $\Delta\varphi_{nR} = \varphi_{n} - \varphi_{R}$.
The FT of $I_\mathrm{RS}$, i.e. $F_\mathrm{RS}$, can be conveniently used to extract the atomic distribution
in the lattice. Apart from an overall constant factor $\lambda f$, the spatial frequencies $\nu_{nx}$ and $\nu_{ny}$
directly correspond to the coordinates $x_n$ and $y_n$ of each particle $n$.
\begin{eqnarray}
\nu_{nx}  = \frac{x_n}{\lambda f} + \frac{\sin\theta \cos\phi }{\lambda}, \nonumber \\
\nu_{ny}  = \frac{y_n}{\lambda f} + \frac{\sin\theta \sin\phi}{\lambda}
\label{eq:nunR}
\end{eqnarray}
The offsets, $\sin\theta \cos\phi / \lambda$ and $\sin\theta \sin\phi /\lambda$, can be tuned by adjusting the direction of
the incident reference beam (see Eq.~(\ref{eq:k_R})). As in spatial heterodyne detection, they are used to shift the peaks of the signal $F_\mathrm{RS}$ away from
the origin to separate them from the peaks of $F_\mathrm{0}$ and $F_\mathrm{S}$. Resolving Eq.~(\ref{eq:nunR}) for the atomic
coordinates $x_n$ and $y_n$ yields
\begin{eqnarray}
x_n & = \lambda f \nu_{nx} - f \sin\theta \cos\phi , \nonumber \\
y_n & = \lambda f \nu_{ny} - f \sin\theta \sin\phi .
\label{eq:xnyn}
\end{eqnarray}

\section{Numerical Calculations}
\label{simulations}
In this section, we present the results of our numerical calculations. First, we specify the used parameters and discuss the
case of a noiseless detection. Then, we include detection noise and analyze its influence on the reconstruction fidelity. Finally, we compare
our method with direct fluorescence detection and estimate its sensitivity to mechanical vibrations.

\subsection{Parameters and details}

In the following, we consider an ensemble of $N_\mathrm{A}$ = 50 $^6\textrm{Li}$ atoms in a 2D lattice
(see Fig.~\ref{fig:lattice}).
The wavelength of the coherent probe and reference laser beams is set to $\lambda$ = 671\,nm,
close to the $D_2$ transition of $^6\textrm{Li}$. The saturation intensity is $I_\mathrm{sat}=2.54$~mW/cm$^{2}$ at a natural linewidth of
$\Gamma=2\pi\times 5.87$~MHz.
The focal length of the collimation lens is chosen to be $f$ = 7\,mm and the
numerical aperture (NA) is 0.71, which matches typical parameters of a custom long working distance objective
(see Fig. \ref{fig:scheme0}).

In the given case, we set the reference beam waist to $w=5~mm$ (see Eq. \ref{eq:gR}).
The  illuminated area in the detection plane, which is located 70\,mm away
from the lens, has a radius of about 7\,mm. In our simulations we consider only a part of this area, namely
a square section of 10$\times$10\,mm$^2$. The CCD pixel size is assumed to be
$A_\mathrm{P}$ = $7\times 7\,\mu$m$^2$, the quantum efficiency is set to $Q=0.8$.
We choose an acquisition time $T_\mathrm{ac}$ of 200\,$\mu$s. On the considered time scale mechanical vibrations
and particle tunneling inside the lattice can be neglected.

Two fundamental parameters are varied: The average number of photons $N_\mathrm{ph}$ scattered by a single
atom into the entire solid angle $4\pi$ within $T_\mathrm{ac}$, and the total power $P_\mathrm{R}$ of the reference beam. In
the present study, we consider the ranges $100 \leq N_\mathrm{ph} \leq 500$ and
$10^{-8}\,\mathrm{W} \leq P_\mathrm{R} \leq 10^{-2}\,\mathrm{W}$.
 Given an average number of scattered photons $N_\mathrm{ph}$,
the corresponding electric field strength $E_\mathrm{A0}$ is obtained from Eq.~(\ref{eq:Nph2EA0}). Using Eqs.~(\ref{eq:Rs}) and
(\ref{eq:Icoh}), we verify that with these parameters we stay in the regime of mainly coherent emission.

\subsection{Calculating the intensity pattern}
We calculate the image captured by the CCD camera as follows. First, the intensity profile $I_\mathrm{D}(x,y)$ in the considered
section of the detection plane is calculated using Eq.~(\ref{eq:TotalField}). Then, the intensity $I_\mathrm{CCD}(x_p,y_p)$
collected by a CCD pixel at position $(x_p,y_p)$ is obtained by averaging over all intensity contributions covered by the
corresponding pixel area. In contrast to $x$ and $y$, the coordinates $x_p$ and $y_p$ exhibit only discrete values.
$I_\mathrm{CCD}$ is converted into an integer number $N_\mathrm{D}$ of nominally incident photons, ignoring for now
photon shot noise,
\begin{equation}
N_\mathrm{D}(x_p,y_p) = \mathrm{round}\left(\frac{I_\mathrm{CCD}(x_p,y_p) T_\mathrm{ac} A_\mathrm{P}}{\hbar \omega}\right).
\label{eq:I2ND}
\end{equation}
The  output signal of a CCD camera in counts is
\begin{equation}
N_\mathrm{counts}(x_p,y_p) = \mathrm{round} \left(\frac{N_D(x_p,y_p) Q}{G}\right).
\label{eq:counts}
\end{equation}
$G$ is the gain, i.e. the number of accumulated electrons that correspond to one count.
In the following we use $G = 1$ for our calculations. We note that values up to  $G = 10$ yield almost the same
results as for $G=1$.

\subsection{Calculations without noise}

Let us start the discussion of our calculations by considering  the idealized situation of absent noise. Furthermore, for
the purpose of better illustration, we choose an example where the power of the reference laser is comparatively low
($P_\mathrm{R}$ = $10^{-8}$\,W). For this choice, interference fringes are clearly visible, since the ratio $I_{RS}/I_0$ is
comparatively high.

\begin{figure}[!h]
\includegraphics[scale=1.0]{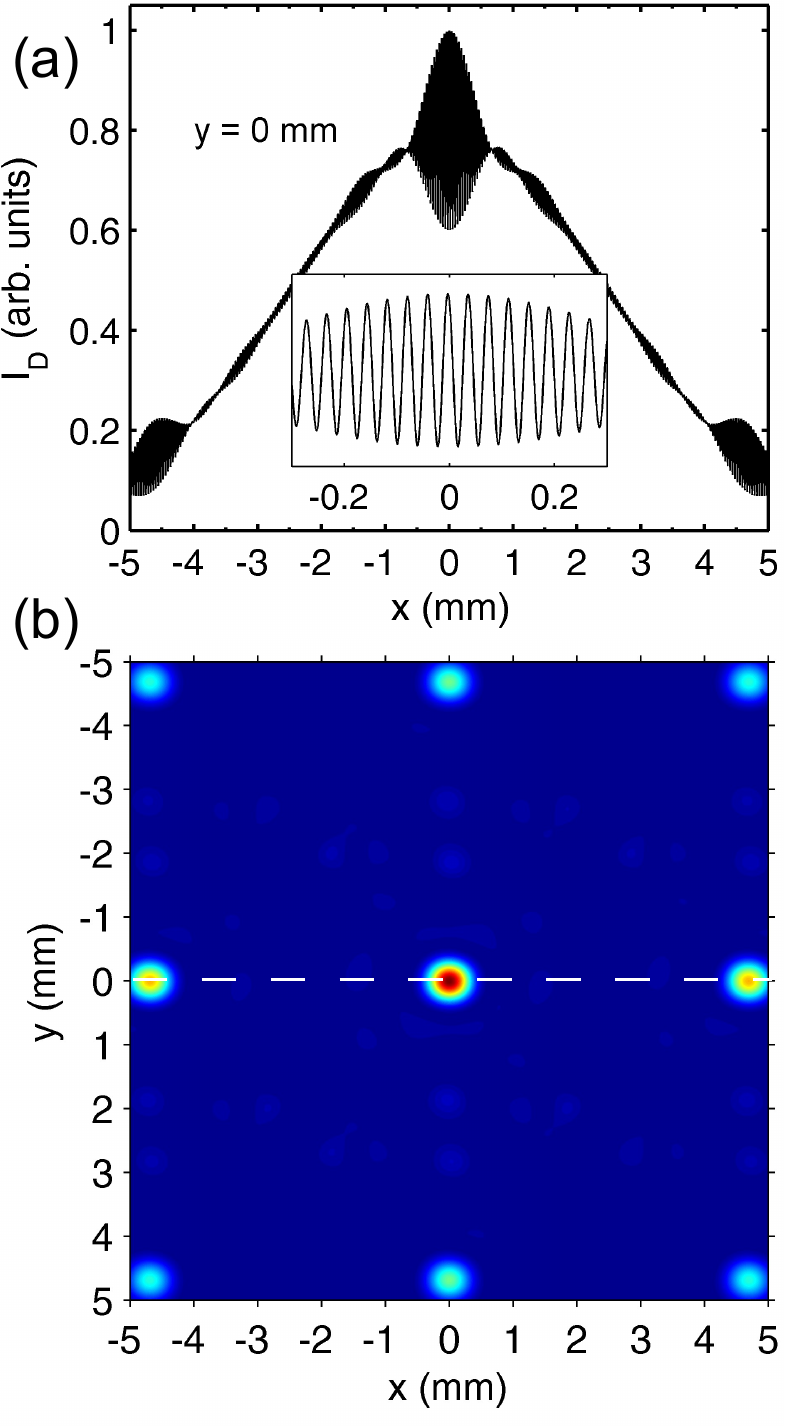}
\caption{(a) 1D cut through the calculated 2D intensity profile $I_\mathrm{D}(x,y)$ in the detection plane along the x axis for
y = 0. The intensity profile results from a superposition of the scattered probe beam and the broad Gaussian reference beam.
An enlargement of the central part (inset) clearly reveals a sinusoidal interference pattern. For illustration purposes, a very
low reference signal has been used in this model calculation, such that the interference fringes are clearly visible on the
Gaussian background signal. (b) 2D intensity distribution in the detection plane without reference beam (false color image; blue: low, red: high intensity). It strongly resembles
the diffraction pattern of a 2D square lattice where the zeroth and first-order peaks are located in the center and at the edges,
respectively.}
\label{fig:scheme}
\end{figure}

Figure~\ref{fig:scheme}a shows a cut through the corresponding intensity profile $I_\mathrm{D}(x,y)$ along the x axis at y = 0
calculated with $N_\mathrm{ph}$ = 500.
The pronounced oscillations on top and at the tails of a Gaussian profile as well as weak oscillations in between arise from the
interference between the scattered probe light and the Gaussian-shaped reference beam. Since the relevant information about the
atom positions is stored in these interference fringes, the period length of the oscillations must be large enough to be resolved
even after averaging intensity values within a pixel (see explanation above).
We achieve this by using a small angle of incidence $\theta \approx 1^\circ$.
 This results in a sufficiently large
period length of about 40\,$\mu$m as revealed by the inset of Fig.~\ref{fig:scheme}a. The angle corresponds to a distance between
the focus of the reference beam and the atomic ensemble of about 100\,$\mu$m (see Fig.~\ref{fig:scheme0}). 

The emergence of the
pronounced interference peaks at $0, \pm 4.5$mm in Fig.~\ref{fig:scheme}a can be understood as follows. To first order, the light
scattered by the rectangular array of atoms  resembles the diffraction pattern of a perfect 2D square lattice, as depicted in
Fig.~\ref{fig:scheme}b. The quickly-oscillating intensity peaks in the center and at the edges in Fig.~\ref{fig:scheme}a are the
corresponding zeroth and first-order diffraction peaks which interfere with the reference beam. The atomic array, however, is not
perfect as a number of lattice sites are unoccupied. As a consequence, the intensity in between the major diffraction
peaks is non-zero. This leads to the weak, but still clearly visible interference patterns in Fig.~\ref{fig:scheme}a between the
strong oscillations in the middle and at the edges. The information about occupied lattice sites is contained in these oscillations. In order to resolve them, especially
for a higher reference laser power, the CCD camera needs a large dynamic range (12 bit or better).

\begin{figure}[ht]
\includegraphics{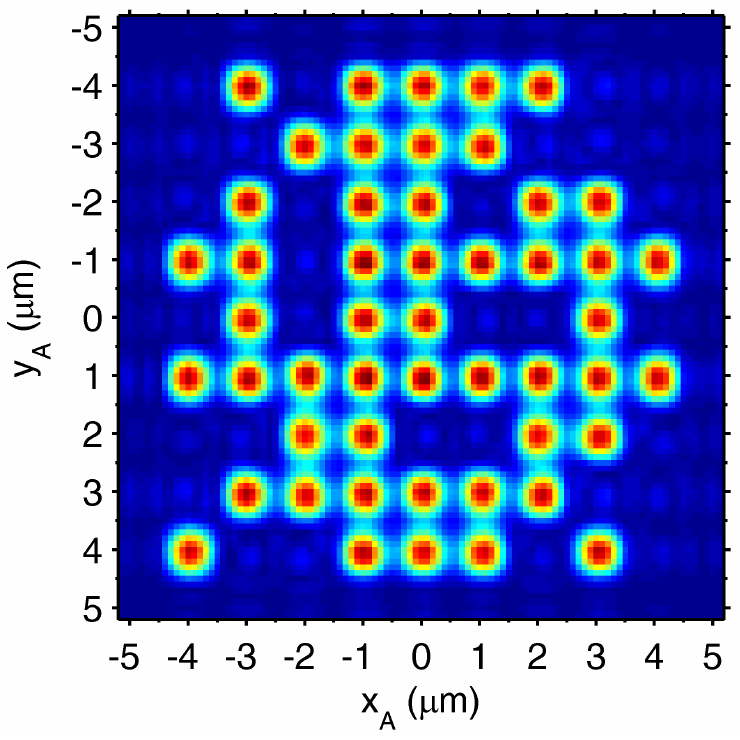}
\caption{ Section of the 2D FT yielding a perfectly reconstructed image of the atomic distribution in Fig.~\ref{fig:lattice}. The false
color plot displays absolute values of the Fourier coefficients (blue: low, red: high FT amplitude). The simulation was performed without noise using the parameters
$N_\mathrm{ph}$ = 150 and $P_R = 10^{-5}$W. }
\label{fig:Back}
\end{figure}

As explained in Sec.~\ref{theoretical_description}, the atomic positions in the lattice can be directly derived from a 2D FT
of the intensity profile $I_\mathrm{D}(x,y)$, or more precisely from a FT of $N_\mathrm{counts}(x_p,y_p)$. An appropriately chosen
section of such a FT is shown in Fig.~\ref{fig:Back}, where the absolute values of the Fourier coefficients are displayed as a
false color image.
The coordinates $x_\mathrm{A}$ and $y_\mathrm{A}$ give the position within the atomic layer and are related to the spatial
frequencies $\nu_x$ and $\nu_y$ of the FT  by (see Eq.~(\ref{eq:xnyn}))
\begin{eqnarray}
x_\mathrm{A} & = \lambda f \nu_x - f \sin\theta \cos\phi, \nonumber \\
y_\mathrm{A} & = \lambda f \nu_y - f \sin\theta \sin\phi.
\label{eq:xAyA}
\end{eqnarray}
The maxima in the 2D plot at $y_\mathrm{A}=0$ correspond to the group of peaks labeled by $F_\mathrm{RS}$ in the schematic 1D
illustration of Fig.~\ref{fig:FFTschematic}. In contrast to Fig.~\ref{fig:scheme}a, $N_\mathrm{counts}(x_p,y_p)$ is calculated
using $N_\mathrm{ph}$ = 150 and  $P_\mathrm{R}$ = $10^{-5}$\,W. A comparison with Fig.~\ref{fig:lattice} reveals that the atomic
distribution is perfectly reconstructed.

\subsection{Speckle and shot noise}

Let us now turn to the realistic situation where the image acquired by the CCD camera is disturbed by different kinds of noise.
These need to be taken into account to understand where the limits of the presented holographic detection method lie. In general,
in an experiment there are several sources which decrease the fidelity of a detection. For a CCD camera, there are photon shot
noise, read-out noise, and dark counts which have to be taken into account.
However, for the case of a relatively strong reference beam and thus of a high light intensity, the dominant detection
noise is given by shot noise.
Shot noise describes fluctuations in the number of detected photons and obeys a Poisson distribution. It is taken into account by replacing $N_D$, calculated from
$I_\mathrm{CCD}(x_p,y_p)$ in Eq.~(\ref{eq:I2ND}), by a Poisson-distributed variable with expectation value $N_D$.

Imperfections of the ideally Gaussian intensity profiles of the probe and reference beams will also have an effect on the
reconstruction. Such corrugations can be caused by effects such as shortcomings in the quality of optical elements or weak stray
reflections of the laser beams. The resulting intensity distribution typically shows high-frequency intensity fluctuations similar
to laser speckle \cite{Da84,Go76}. 
If we assume the fluctuations to occur on a length scale of about 1\,$\mu$m in the detection plane, this kind of noise
adds the intensity $I_\mathrm{SP}(x,y)$ to $I_\mathrm{D}(x,y)$. It is known \cite{Da84,Go76} that this added noise has an
exponentially decreasing probability as a function of $|I_\mathrm{SP}|$
\begin{equation}
 P_\mathrm{SP}(I_\mathrm{SP})\propto \exp\left(-\frac{|I_\mathrm{SP}|}{\alpha I_\mathrm{D}}\right).
\end{equation}
From our own laboratory experience we estimate that the typical amplitude of these fluctuations is on the level of about one
percent. Therefore, we set $\alpha = 0.01$.

In order to take into account read-out noise, we add an integer number $\Delta N_\mathrm{counts}(x_p,y_p)$ to
$N_\mathrm{counts}(x_p,y_p)$. This noise is obtained from a zero-centered normal distribution with a standard deviation of 3 counts (typical specification of a commercial electron multiplying CCD camera).

\begin{figure}[!h]
\includegraphics[scale=1.0]{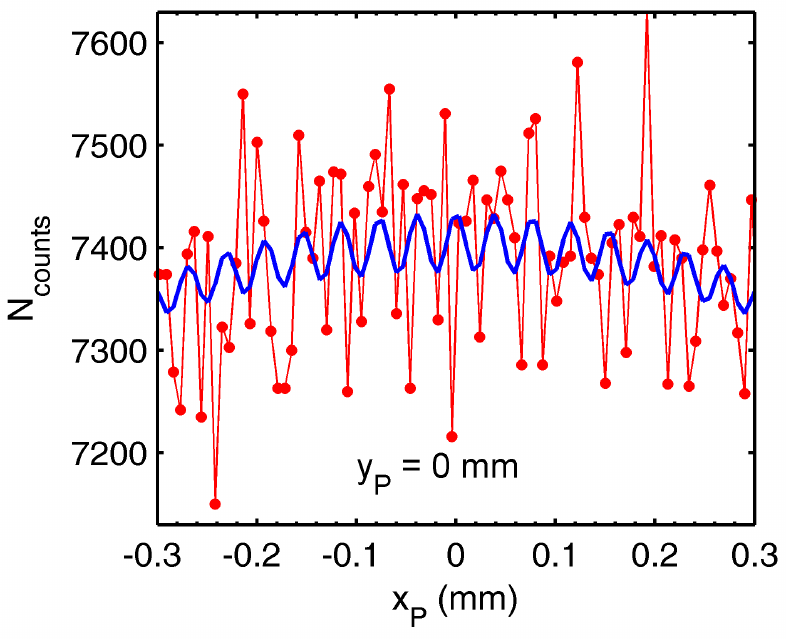}
\caption{1D cut through the calculated CCD image $N_\mathrm{counts}(x_p,y_p)$ along the $x_p$ axis at $y_p$ = 0 including noise (red line/dots).
The parameters used in the calculation are $N_\mathrm{ph}$ = 150 and $P_R = 10^{-5}$W. The blue line depicts the undisturbed interference signal for comparison.}
\label{fig:combined_effects}
\end{figure}

\begin{figure}[ht]
  \includegraphics{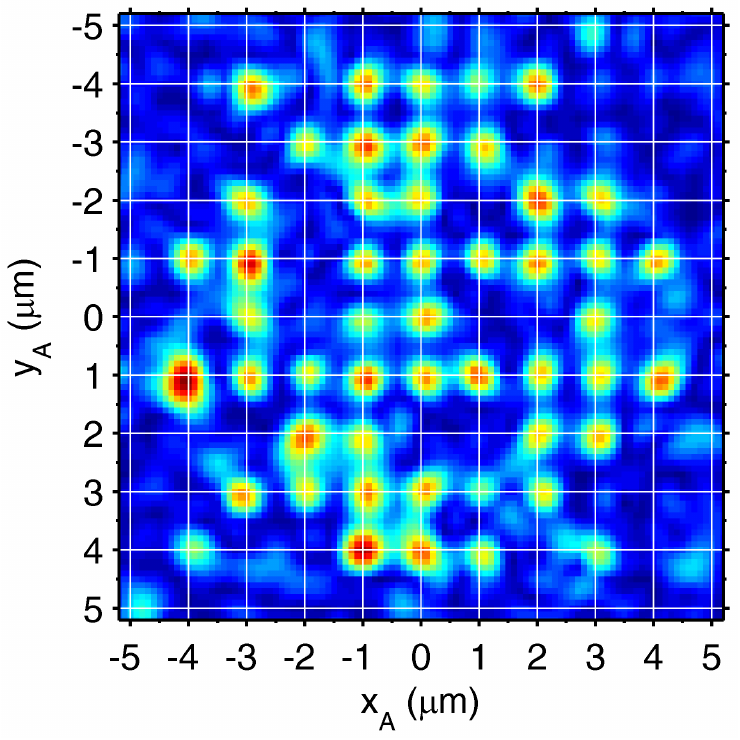}
\caption{Example of a reconstructed image of the atomic distribution (2D FT of $N_\mathrm{counts}(x_p,y_p))$ taking into account speckle, shot,
and read-out noise (blue: low, red: high FT amplitude). The simulation was performed using the parameters $N_\mathrm{ph}$ = 150 and $P_R = 10^{-5}$W. For this example our simple
recognition algorithm (see text) yields a fidelity of 99.2\% to identify the occupation of an individual site.}
\label{fig:Back2}
\end{figure}

The combined effects of intensity averaging, noise, as well as photon counting (see Eq.~(\ref{eq:counts})) are
illustrated in Fig.~\ref{fig:combined_effects}. It depicts a 1D cut through the CCD image $N_\mathrm{counts}(x_p,y_p)$ along the
$x_p$ axis for $-0.3\,{\rm mm} \leq x_p \leq 0.3\,{\rm mm}$ and $y_p$ = 0, calculated with $N_\mathrm{ph}$ = 150 and
$P_R = 10^{-5}$W. In contrast to the inset of Fig.~\ref{fig:scheme}a, which displays the same $x$ range, the interference pattern
is now barely perceptible.
The corresponding 2D FT is shown in Fig.~\ref{fig:Back2}.
In contrast to Fig.~\ref{fig:Back}, it is very  noisy. However, we can still reconstruct the atomic distribution with
a sufficiently high fidelity. 

For this, we use the following simple algorithm.
 We normalize the reconstruction signal (absolute values of the Fourier coefficients) within the FT section
depicted in Fig.~\ref{fig:Back} and Fig.~\ref{fig:Back2}. Next, we place the lattice grid on top as shown in Fig.~\ref{fig:Back2}.
The normalized value at each grid point is compared to a threshold value. If the value lies (below) above the threshold, the
lattice site is identified as (un)occupied. We define a fidelity as the percentage of correctly identified sites.
An analysis of a variety of atomic arrays with different filling factors shows that for the investigated range of parameters $N_\mathrm{ph}$ and
$P_\mathrm{R}$ a threshold value of 0.4 yields the highest fidelity.

\begin{figure}[ht]
\includegraphics{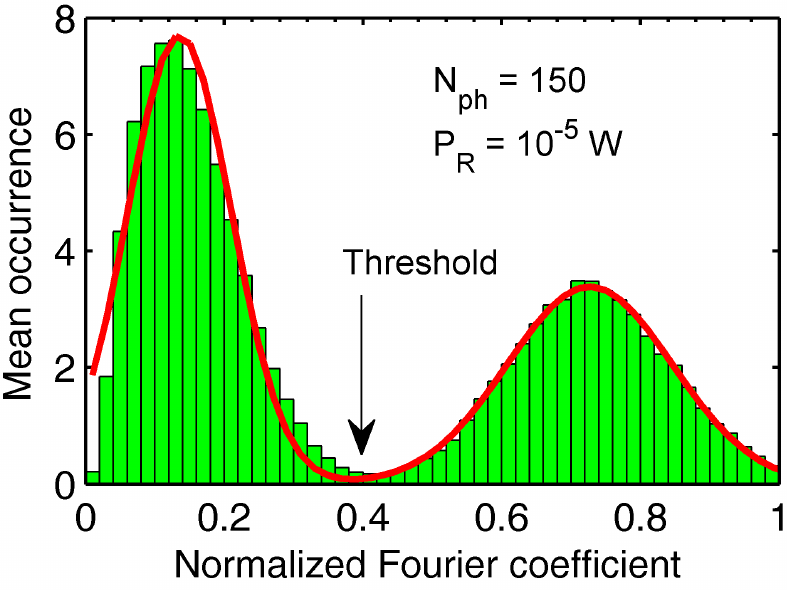}
\caption{ Probability distribution of the normalized Fourier coefficients determined at the optical lattice sites
(see Fig.~\ref{fig:Back2}). The histogram represents an average over 1000 probability distributions obtained for
$N_\mathrm{ph}$ = 150, $P_R = 10^{-5}$W, fixed speckle noise, and randomly varying shot noise. The dip at 0.4 coincides with
the threshold value with highest reconstruction fidelity. The red solid curve are two partially overlapping Gaussians which are
fitted to the histogram.}
\label{fig:histogram}
\end{figure}

\begin{figure}[ht]
\centering
\includegraphics[scale = 1.1]{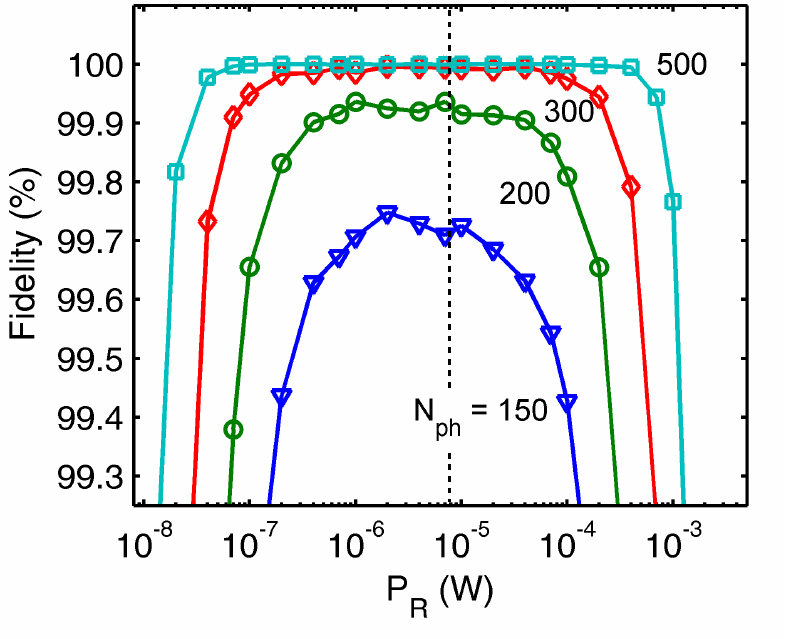}
\caption{Fidelity as a function of $P_R$ for different average total numbers of scattered photons per atom $N_\mathrm{ph}$. Each data
point is obtained by averaging over the fidelities of 1000 reconstructed images (with fixed particle distribution),
calculated with randomly varying shot and read-out noise. Above a critical value of $P_\mathrm{R}$, marked by the dashed line, the pixels near the
center of a CCD camera with a dynamic range of 16 bit start to saturate.}
\label{fig:fidelity}
\end{figure}

The histogram in Fig.~\ref{fig:histogram} displays the
probability distribution of the normalized Fourier coefficients for $N_\mathrm{ph}$ = 150, $P_R = 10^{-5}$W.
 It is obtained by averaging over the probability distributions of 1000 reconstructed images 
 of the particle distribution of Fig. \ref{fig:lattice}. 
 The calculation includes a 
fixed speckle noise and randomly varying shot and read-out noise.
 As shown by the red line in Fig. \ref{fig:fidelity}, the distribution resembles two overlapping Gaussians with a pronounced minimum at 0.4.

In Fig.~\ref{fig:fidelity}, the fidelity is plotted as a function of $P_\mathrm{R}$ for different values of
$N_\mathrm{ph}$. Each data point is again obtained by averaging over the fidelities of 1000 reconstructed images, calculated with
randomly varying shot and read-out noise. For $N_\mathrm{ph} \geq 150$ the fidelity reaches maximum values clearly exceeding 99.5\%, for
$N_\mathrm{ph}$ = 100 (not shown) it is still nearly 98.5\%. 

Depending on the range of $P_\mathrm{R}$, the fidelity is limited by
different kinds of noise. At low and high reference power, read-out noise and speckle noise prevail, respectively. In both cases,
the noise leads to a strong decline of the fidelity. In between, shot noise is dominant. The dependence of the fidelity on
$P_\mathrm{R}$ and $N_\mathrm{ph}$ can be understood by the signal-to-noise ratio (SNR) of the interference fringes on
the CCD camera. Neglecting atomic contributions to $I_\mathrm{0}$ in Eq.~(\ref{eq:I_0}), a rough estimate yields
\begin{equation}
\mbox{SNR} \approx \frac{I_\mathrm{RS}}{\sqrt{I_\mathrm{0}}} \propto E_\mathrm{A0} \propto \sqrt{N_\mathrm{ph}}.
\label{eq:SNR}
\end{equation}
In the fraction $I_\mathrm{RS}/\sqrt{I_\mathrm{0}}$, the field amplitude $E_\mathrm{R0}$ of the reference beam
drops out and the SNR is independent of $P_\mathrm{R}$. As a consequence, the fidelity features a plateau. The width of the plateau as
well as the maximum fidelity decreases with decreasing $N_\mathrm{ph}$. This can be explained by the proportionality of SNR to the
atomic field amplitude $E_\mathrm{A0}\propto \sqrt{N_\mathrm{ph}} $. Above a critical value of $P_\mathrm{R}$, marked by the
dashed line in Fig.~\ref{fig:fidelity}, the pixels near the center of the CCD camera saturate (assuming a dynamic range of 16 bit).
Therefore, in practice the speckle-induced drop should be irrelevant.
\subsection{Comparison with fluorescence detection}
As$\,$ demonstrated$\,$ in $\,$Fig.~\ref{fig:fidelity}, the proposed$\,$ detection$\,$ scheme should yield fidelities higher than 99.5\% even for
moderate numbers of scattered photons. This is achieved by means of the reference beam which amplifies the atom$\-$ic diffraction signal.
In contrast, the direct fluorescence detection method, e.g. used in \cite{Bak09,She10,Zw15,Kuh15,MG15}, does not involve
such a reference beam. 
During detection several thousands of photons are scattered by
a single atom. As a disadvantageous consequence, the atoms are strongly heated and may hop between lattice sites 
even in the case of deep optical lattices (see e.g. \cite{MG15}). Therefore, complex cooling techniques have to be applied.

To compare our scheme with the fluorescence detection, we estimate the particle heating.
We assume that the particles are initially in the vibrational ground
state $\ket{v=0}$ of a deep optical lattice with a depth of 2.5\,mK and a Lamb-Dicke parameter of  $\eta=0.23$ (see \cite{Gro15}).
During detection, the particles scatter 150 photons per atom. The transition probability from $\ket{v=n}$ to $\ket{v=n\pm1}$ for a single
scattering event is given by $\eta^2\left(n+1\right)$ and $\eta^2n$, respectively. An estimate based on random walk yields that 99\% of the atoms end up at a vibrational
state $\ket{v_{Final}\leq 24}$. The excitation to higher vibrational states reduces the tunneling time of a particle inside the lattice. However,
since the tunneling time of a particle in state $\ket{v_{Final}=24}$ is on the order of 1~ms, i.e. long compared to the acquisition time
$T_\mathrm{ac} = 200\,\mu$s, tunneling can be neglected. This means that the heating due to light scattering should hardly influence the
reconstruction fidelity. Therefore, our scheme might open the path to circumvent additional cooling during detection.

\subsection{Mechanical vibrations}

In terms of a technical issue of the proposed scheme, we need to take into account the sensitivity of the setup to mechanical
vibrations. For this, we consider Eq.~(\ref{eq:I_RP}) and Eq.~(\ref{eq:xAyA}). During the acquisition time, the relative phases
$\Delta\varphi_{nR}$ between reference and scattered laser fields may vary, leading to a blurring of the contrast of the
interference fringes. 
A jitter $\delta\theta$ in the reference angle $\theta$ leads to a similar effect. In order to
estimate the influence of the jitter, we rewrite $x_\mathrm{A}$ in Eq.~(\ref{eq:xAyA}) for angles close to $\theta \approx 1^\circ$
(as used in our simulations) with $\phi = 45^\circ$ fixed:
\begin{equation}
x_\mathrm{A} = \lambda f \nu_x - \frac{f}{\sqrt{2}} \theta.
\label{eq:xAyAsmall}
\end{equation}
A jitter $\delta\theta$ thus causes a blurring $\delta x = f \delta\theta/\sqrt{2}$ of the coordinates in the reconstruction.
If we demand $\delta x \ll a$, the jitter has to be much smaller than
$\sqrt{2}\times 1\,\mu\mathrm{m}/f \approx 200 \,\mu$rad. This should not pose a problem since pointing stabilities of $10~\mu$rad
or better are typical in an optical lab environment. Furthermore, achieving fluctuations in the relative phase
$\Delta\varphi_{nR} \ll \pi$ is standard on an optical table.

\section{Conclusion}
\label{conclusion}

In conclusion, we propose a holographic scheme for site-resolved detection of a 2D gas of ultracold atoms in an optical lattice.
We have discussed the method for the example of 50 lithium atoms in a square optical
lattice, but it will also work for larger sample sizes, other atomic elements, or other lattice geometries.
The method features a high detection fidelity ($>99.5\%$) even for a low number of scattered photons per atom ($\approx 150$) in the presence of detection noise.

The low number of scattered photons might open the path for single site detection without additional cooling.
Moreover it might allow for imaging multiple occupancy of a single lattice site.

\begin{acknowledgement}
JHD would like to thank J\"{u}rgen Eschner for an interesting discussion on photon scattering.
The authors thank all the members of the Institut f\"{u}r Quantenmaterie.
This work was supported by the German Research Foundation DFG within the SFB/TRR21.
\end{acknowledgement}

\end{document}